# *Attention-based Saliency Maps Improve Interpretability of Pneumothorax Classification*

***Title of the manuscript:*** Attention-based Saliency Maps Improve Interpretability of Pneumothorax Classification

**Authors:**

- Alessandro Wollek, M.Sc.*[1,2],
- Robert Graf, M.Sc. *[1,2]
- Saša Čečatka, MD. [3]
- Nicola Fink, MD. [3]
- Theresa Willem, M.A. [4]
- Bastian O. Sabel, MD. [3]
- Dr. Tobias Lasser [1,2]

[*] Alessandro Wollek and Robert Graf contributed equally to this work.

[1] Munich Institute of Biomedical Engineering, Technical University of Munich, Boltzmannstr. 11, Garching b. München, 85748, Bavaria, Germany

[2] Department of Informatics, Technical University of Munich, Boltzmannstr. 3, Garching b. München, 85748, Bavaria, Germany

[3] Department of Radiology, University Hospital LMU, Marchioninistrasse 15, München, 81377, Bavaria, Germany

[4] Munich School of Technology in Society, Technical University of Munich, Arcisstraße 21, München, 80333, Bavaria, Germany

**Work originated from:** Munich Institute of Biomedical Engineering, Technical University of Munich, Boltzmannstr. 11, Garching b. München, 85748, Bavaria, Germany.

**Corresponding Author:**
Alessandro Wollek
- Phone: +49 89 289 10840
- Email: alessandro.wollek@tum.de
- Address: Boltzmannstr. 11, Garching b. München, 85748, Bavaria, Germany

**Funding:**
The research for this article received funding from the German federal ministry of health's program for digital innovations for the improvement of patient-centered care in healthcare [grant agreement no. 2520DAT920].

**Data Sharing Agreement:**
*All data sets used are publicly accessible:*
- *CheXpert:* https://stanfordmlgroup.github.io/competitions/chexpert/
- *Chest X-Ray 14:* https://www.kaggle.com/datasets/nih-chest-xrays/data
- *MIMIC CXR:* https://physionet.org/content/mimic-cxr-jpg/2.0.0/
- VinBigData: https://www.kaggle.com/c/vinbigdata-chest-xray-abnormalities-detection/
- SIIM ACR: https://www.kaggle.com/c/siim-acr-pneumothorax-segmentation
- Code: https://gitlab.lrz.de/IP/publishing/attention-based-saliency-map

**Summary statement:**
**Vision transformers performed comparably with convolutional neural networks in pneumothorax classification on chest radiographs, and their resulting attention-based saliency maps had greater interpretability than gradient-weighted class activation maps.**

**Key Points:**
- Vision transformers (ViT) achieved similar performance to conventional convolutional neural networks (CNN) for pneumothorax classification on chest radiographs on three data sets (the area under the receiver operating characteristic curve (AUC) ranged from 0.84 to 0.95 for ViT, and AUCs ranged from 0.83 to 0.87 for CNN).
- Attention-based saliency maps were considered useful in more cases (47 %) than gradient-weighted class activation-mapping (GradCAM) (39 %) for pneumothorax diagnosis by radiologists.

- Attention-based saliency maps explained model decisions more effectively than GradCAM on all three data sets and across all evaluation metrics (perturbation test, sensitivity-n, effective heat ratio, intra-architecture repeatability, and interarchitecture reproducibility).

**Abbreviations:**

| | |
|---|---|
| **AI** | artificial intelligence |
| **CNN** | convolutional neural network |
| **ViT** | vision transformers |
| **CXR** | chest radiograph |
| **TMME** | Transformer-Multi-Modal-Explainability |
| **GradCAM** | gradient-weighted class activation mapping |
| **AUC** | area under the receiver operating characteristic curve |

# Abstract

**Purpose:** To investigate chest radiograph (CXR) classification performance of vision transformers (ViT) and interpretability of attention-based saliency using the example of pneumothorax classification.

**Materials and Methods:** In this retrospective study, ViTs were fine-tuned for lung disease classification using four public data sets: CheXpert, Chest X-Ray 14, MIMIC CXR, and VinBigData. Saliency maps were generated using transformer multi-modal explainability and gradient-weighted class activation mapping (GradCAM). Classification performance was evaluated on the Chest X-Ray 14, VinBigData, and SIIM-ACR data sets using the area under the receiver operating characteristic curve analysis (AUC) and compared with convolutional neural networks (CNNs). The explainability methods were evaluated with positive/negative perturbation, sensitivity-n, effective heat ratio, intra-architecture repeatability and interarchitecture reproducibility. In the user study, three radiologists classified 160 CXRs with/without saliency maps for pneumothorax and rated their usefulness.

**Results:** ViTs had comparable CXR classification AUCs compared with state-of-the-art CNNs 0.95 (95 % CI: 0.943, 0.950) vs. 0.83 (95 %, CI 0.826, 0.842) on Chest X-Ray 14, 0.84 (95 % CI: 0.769, 0.912) vs. 0.83 (95 % CI: 0.760, 0.895) on VinBigData, and 0.85 (95 % CI: 0.847, 0.861) vs. 0.87 (95 % CI: 0.868, 0.882) on SIIM ACR. Both saliency map methods unveiled a strong bias towards pneumothorax tubes in the models. Radiologists found 47 % of the attention-based saliency maps useful and 39 % of GradCAM. The attention-based methods outperformed GradCAM on all metrics.

**Conclusion:** ViTs performed similarly to CNNs in CXR classification, and their attention-based saliency maps were more useful to radiologists and outperformed GradCAM.

# Introduction

Artificial Intelligence (AI) has the potential to improve medical processes (1,2) by increasing human performance and speed, but its black-box nature leads to big hurdles for decision-making support (3,4). Radiologists can confirm or reject AI-based identification of an important finding, such as a potentially life-threatening pneumothorax, most optimally when the location is provided; evaluation becomes more difficult and time-consuming when the area in question remains to be located (5). Even when deep learning models perform similarly to radiologists (6,7), erroneous decisions made by the models differ in nature from those made by human experts (7). Therefore, radiologists working with such a model need appropriate methods to understand its strengths and limitations.

A saliency map addresses this problem by highlighting locations that are relevant to the model's decision-making process (**Figure 1**). Predictions based on insufficient evidence, such as a chest tube as sole evidence for pneumothorax, allow the radiologist to confidently reject the model's prediction.

Current state-of-the-art image classifiers are based on the vision transformer (ViT)(8,9), Compared with convolutional neural networks, ViTs leverage the attention mechanism that is structurally easier to interpret, as they capture the importance of spatial token relationships (e.g., image patches) to model predictions. We created attention-based saliency maps using transformer multi-modal explainability (TMME)(10). TMME is an extension of the roll-out mechanism, the matrix multiplication of all attention maps. The attention heads are merged by taking the minimum or mean. Skip-connections of attention layers are accounted for by adding an identity matrix to the merged attention matrix. Additionally, all heads are multiplied with their derivatives conditioned on a chosen class. Negative values are set to zero before applying roll-out. As baseline, we applied the frequently used gradient-weighted class activation mapping (GradCAM) technique (11).

To date, there has been limited research on the use of saliency maps for medical purposes (12–18), which all focus on comparing saliency maps and segmentation rather than explaining the basis for a model's decision. There is a need to investigate the reliability of saliency maps on chest radiographs (CXRs) and their usefulness to radiologists. To the best of our knowledge, medical studies have not yet been conducted on attention-based explainers.

The aim of this exploratory study was to evaluate the feasibility of ViTs for CXR classification and interpretability of attention-based saliency maps for clinical decision support using the example of pneumothorax classification.

# Materials and Methods

## Data Sets

This retrospective study used five public CXR data sets: CheXpert (19), Chest X-Ray 14 (20), MIMIC CXR (21), VinBigData (22), and Society for Imaging Informatics in Medicine-American College of Radiology (SIIM-ACR) (23) (**Figure 2**). CheXpert consists of 224,316 CXR of 65,240 patients, Chest X-Ray 14 of 112,120 CXR of 30,805 patients, MIMIC CXR of 377,110 CXR of 65,379 patients, VinBigData of 15,000 labeled CXR, and SIIM-ACR of 2669 pneumothorax CXR. Institutional review board review and patient informed consent were not required for this study. We trained on CheXpert, Chest X-Ray 14, MIMIC CXR and VinBigData, validated on the holdout set of CheXpert, and tested on the data sets containing image annotations: the holdout set of Chest X-Ray 14, 7500/15000 (50%) of the VinBigData set, and all SIIM-ACR data. All sets were split without patient overlap and are publicly accessible:

CheXpert: https://stanfordmlgroup.github.io/competitions/chexpert/

Chest X-Ray 14: https://www.kaggle.com/datasets/nih-chest-xrays/data

MIMIC CXR: https://physionet.org/content/mimic-cxr-jpg/2.0.0/

VinBigData: https://www.kaggle.com/c/vinbigdata-chest-xray-abnormalities-detection/

SIIM-ACR: https://www.kaggle.com/c/siim-acr-pneumothorax-segmentation

## Model Training and Development

We fine-tuned the ViT variant 'deit_base_distilled_patch16_224'(9) on the overlapping training data set classes: pneumothorax, cardiomegaly, consolidation, pleural effusion and atelectasis (Figure 2). We replaced the final softmax layer with a sigmoid. We used stochastic gradient descent (SGD) with a learning rate of 0.001 and a batch size of 64 and trained for 500 epochs with early stopping. We randomly oversampled the images. During training, we applied random affine translations (-15 ° – 15 °, translate 0.05, scale 0.9 – 1.05) and random horizontal flip. All images were normalized according to the ImageNet mean and standard deviation and resized to 224 x 224 pixels. We compared our network with a pre-trained DenseNet-121 (24), commonly used for CXR classification (6,25). Our model was fine-tuned using the same data and augmentations, SGD with a learning rate of 0.001 and batch size of 32, and trained for 30 epochs with early stopping.

## Saliency Map Evaluation Metrics

***Positive and Negative Perturbation Test.*** We continuously removed patches (*16x16* pixels) of the input image according to their saliency map importance and then generated a new saliency map. We removed the most important patches in the positive perturbation test and least important patches in the negative perturbation test. Model confidence was measured while removing image patches  AUC was calculated. We followed the procedure by Chefer et al. (28), but instead of blackening the removed patches, we replaced them with patches of an image with zero predictive confidence of the network, guaranteeing a convergence to zero instead of random predictions after most pixels were replaced.

***Sensitivity-n.*** The sensitivity-n test (29) applies random masks onto the input image. Masks lowering the confidence should correlate with a good saliency map. We used 200 random masks consisting of *n* random tokens (*16x16* pixels). The Hadamard products of saliency map × mask and the change in confidence were compared with the Pearson product-moment correlation coefficients (30). Scores were averaged, and the test was repeated with different numbers of tokens *n*, logarithmically distributed between 1 and half the total number tokens. We computed the AUC between token number and correlation score. We used the test images that were predicted as a pneumothorax by the ViT.

***Effective Heat Ratio.*** We used the effective heat ratio (EHR) (29) to show that the saliency maps highlight regions that align with prior medical knowledge. The test first produces a binary mask of the saliency map given a threshold. The EHR is the fraction between the threshold area inside the ground truth and the complete threshold area. The thresholds were computed in equidistant steps. We computed the AUC over all EHRs and thresholds.

**Intra-architecture repeatability and interarchitecture reproducibility**. Following Arun et al. (12)**,** we compared the similarity between saliency maps of different models. We compared the mean structural similarity index measure (SSIM) of 1000 saliency maps in the SIIM-ACR dataset on pneumothorax CXRs.

## User Study

To evaluate clinical usefulness, we surveyed one board-certified radiologist with more than ten years' experience (B.S.), one fourth year (N.F.) and one first year radiology resident (S.C.) from Klinikum der Ludwig-Maximilians-Universität München (hereafter referred to as "radiologists"). We compared GradCAM and TMME on images with and without pneumothorax. We had to limit our investigation to pneumothorax detection due to available segmentations and study participation time. We sampled 160 images from the SIIM-ACR data set, 110 with and 50 without pneumothorax. We included 70 TMME, 70 GradCAM, 10 artificial saliency maps based on segmentations and 10 random saliency maps generated by applying Gaussian-blur and multiplying repetitively with Perlin-noise (**Table 1**). We calibrated model predictions using histogram binning (30).

In two consecutive parts, the radiologists assessed the presence of pneumothorax when first given the CXR and model prediction and then when additionally given a saliency map (**Figure 3**). Furthermore, they rated usefulness of the saliency maps on a scale from 1-5 (strongly disagree – strongly agree).

## Statistical Analysis

Model performance was assessed using receiver operating characteristics (ROC), sensitivity, and specificity at maximum F1-score with 95% confidence intervals (calculated using the non-parametric bootstrap method with 10,000-fold resampling at the image level). We compared model classification performance using the area under the ROC curve (AUC). AUC comparison was performed using fast implementation (31) of the non-parametric approach of DeLong et al. (32). Since our analysis is exploratory and involves multiple comparisons P values are provided but significance claims are not made. Radiologist pneumothorax classification performance with and without saliency maps was measured using sensitivity and specificity. Statistical analyses were performed using NumPy (https://numpy.org) version 1.21.5 and SciPy (https://scipy.org) version 1.7.3 in this study.

# Results

## CXR Classification Performance

As the focus of this study lies on pneumothorax, we report the performance results of the other classes, atelectasis, cardiomegaly, consolidation, and pleural effusion, only briefly here. The ROC curves and corresponding AUCs with 95 % CIs are shown in the supplementary materials (**Figure S. 2:** - **Figure S. 5:** EH). For these classes, the ViT outperformed the DenseNet on both the Chest X-Ray 14 and VinBigData data sets. Regarding pneumothorax classification, the ViT achieved a higher AUC than DenseNet on Chest X-Ray 14 (AUCs, 0.95 [95 % CI: 0.94, 0.95] vs. 0.83 [95 % CI: 0.83, 0.84]; p<0.001). We found no evidence of a difference between ViT and DenseNet performance on VinBigData (AUCs, 0.84 [95% CI: 0.77, 0.91] vs. 0.83 [95% CI: 0.76, 0.90]; p = 0.67). The DenseNet pneumothorax classification resulted in a higher AUC than ViT on the SIIM-ACR dataset (AUCs, 0.87 [95% CI: 0.87, 0.88] vs. 0.85 [95% CI: 0.85, 0.86]; p<0.001; **Figure 4**). F1-score, sensitivity, and specificity are reported in **Table 2**.

## Saliency Map Evaluation

The following tests measure how well each saliency method highlights important regions for the ViT's prediction. Across all saliency map tests and data sets, TMME performed better than GradCAM (**Figure 5)**.

On the positive perturbation test, where a lower value is better, as the most important tokens are removed first, TMME and GradCAM measured 0.04 (95% CI: 0.036, 0.044) vs. 0.12 (95% CI: 0.110, 0.135) on SIIM-ACR, 0.03 (95 % CI: 0.030, 0.037) vs. 0.12 (95 % CI: 0.111, 0.137) on Chest X-ray 14, and 0.02 (95 % CI: 0.013, 0.037) vs. 0.09 (95% CI: 0.048, 0.135) on VinBigData, respectively.

The negative perturbation test results, where a higher value is better, as the most important tokens are removed last, were 0.67 (95% CI: 0.65, 0.69) for TMME vs. 0.37 (95% CI: 0.34, 0.39) for GradCAM on SIIM-ACR, 0.55 (95% CI: 0.52, 0.57) vs. 0.25 (95% CI: 0.23, 0.27) on Chest X-ray 14, and 0.73 (95% CI: 0.60-0.85) vs. 0.55 (95% CI: 0.40-0.67) on VinBigData.

Sensitivity-n (higher is better) values for TMME compared with GradCAM were 14 (95 % CI: 12.59, 15.32) vs. 6 (95 % CI: 4.44, 6.57) on SIIM-ACR, 11 (95 % CI: 9.85, 13.14) vs. 4 (95 % CI: 2.83, 5.17) on Chest X-ray 14, and 9 (95 % CI: 1.21, 17.98) vs. 7 (95 % CI: 0.55, 13.00) on VinBigData datasets, respectively.

TMME achieved an EHR (higher is better) of 0.16 (95% CI: 0.142, 0.171) vs. 0.11 (95 % CI: 0.099, 0.122) for GradCAM on SIIM-ACR, 0.26 (95% CI: 0.199, 0.318) vs. 0.14 (95% CI: 0.100, 0.193) on Chest X-ray 14, and 0.33 (95% CI: 0.237, 0.434) vs. 0.22 (95% CI: 0.126, 0.275) on VinBigData datasets.

For intra-architecture repeatability, TMME had an average SSIM score of 0.57 (95% CI: 0.562, 0.578) vs. a GradCAM SSIM score of 0.12 (95% CI: 0.105, 0.126). Comparing different models, TMME had an inter-architecture reproducibility SSIM score of 0.47 (95% CI: 0.465, 0.481) vs. 0.08 (95% CI: 0.074, 0.091) for GradCAM (**Figure S. 6**).

During visual evaluation of the produced saliency maps, we detected that on some images, both GradCAM and TMME saliency maps highlighted confounders, such as a chest tube, instead of pneumothorax (**Figure 6**).

## User Study

Given the frontal CXR and the ViT's prediction alone, the radiologists achieved a sensitivity of 64% (211/330; 95% CI: 59 %, 69 %) and specificity of 84% (126/150; 95% CI: 78 %, 90 %). Sensitivity improved to 65% (216/330; 95% CI: 60 %, 70 %) and specificity remained at 84% (126/150; 95% CI: 78 %, 90 %) after additionally showing the saliency map. Excluding the model's incorrect predictions resulted in a sensitivity of 77% (162/210; 95% CI: 72 %, 83 %) before viewing the saliency map and 79% (165/210; 95% CI: 73 %, 84 %) after; specificity was 93% (56/60; 95% CI: 86 %, 98 %) before and 93% (56/60; 95% CI: 87 %, 99 %) after viewing the saliency map. Showing GradCAM saliency maps did not improve sensitivity or specificity. TMME saliency maps improved sensitivity from 61% (83/135; 95% CI: 53 %, 70 %) to 64% (86/135; 95% CI: 55 %, 71 %) (**Figure S. 7**).

The artificial saliency maps, based on ground truth segmentations, were rated useful in 63 % (19/30) true positive cases and not useful in most false negative cases (26/30, 87 %) (**Table 3**). Both TMME and GradCAM were rated useful in 47/90, 52% and 46/90, 51% true positive predictions, respectively. Overall, TMME was rated useful (99/210, 47%) more often than GradCAM (82/210, 39%), especially for false model predictions (TMME: 41/90, 45% vs. GradCAM: 27/90, 30%).

# Discussion

In this study, we trained a ViT on several CXR data sets to generate attention-based saliency maps to evaluate the performance of attention-based saliency maps compared with GradCAM and their usefulness in assisting radiologists in detecting pneumothoraces. We found that ViTs achieve similar results on CXR classification compared with convolutional neural networks (CNN; AUCs ranged from 0.84 to 0.95 for ViT, and from 0.83 to 0.87 for CNN), despite the limited amount of training data, as ViTs require more training data than CNNs, encouraging further research.

Given more training data we expect the ViTs to significantly outperform CNN-based architectures, as shown in other computer vision tasks (**Figure S. 8**).

Furthermore, we found that attention-based TMME saliency maps outperformed GradCAM across all metrics. This result was further confirmed by a difference in usefulness for pneumothorax detection by radiologists seeing TMME saliency maps compared with GradCAM or no saliency map, suggesting that attention-based saliency maps are more useful in clinical decision support than GradCAM.

We attribute the drastically different results of GradCAM and TMME in some cases, where GradCAM visualization did not allow interpretation, to the inner workings of each method: GradCAM uses a single layer, while TMME combines the attention of multiple layers of the network, making it more robust.

Saliency maps run the risk of human tendency to over-interpret a given visual result (33) and must be evaluated systematically. Model or data biases, such as chest tubes implying a pneumothorax, can be discovered using saliency maps. This phenomenon was observed in the literature before (34) and highlights the importance of increasing the interpretability of such models. These insights can then be used to improve both data quality and architectural choices.

While the results suggest that the radiologists were biased by incorrect model predictions, the saliency map improved the readers' sensitivity and could be used to prevent confirmation bias (35). In a post-study discussion, participants reported that the saliency maps helped them to better assess the strengths and weaknesses of the model, such as the model's focus on the presence of a chest tube in predicting pneumothorax probability. Despite the small number of participants in our study, improved reader sensitivity with attention-based saliency maps encourages further research.

There were several limitations in our study. First, due to the unavailability of segmentation masks for lung diseases besides pneumothorax, the saliency maps could not be analyzed as precisely using the effective heat ratio. In these cases, we used the available but less precise bounding boxes. Second, we believe that model performance, particularly in pneumothorax classification, and the resulting saliency maps were limited by the small amount of available data, even when pooling data from multiple institutions (**Figure S. 8**). ViTs have been shown to substantially outperform CNN-based architectures given enough data (8). Third, while our work indicates that the saliency maps have a positive effect on radiologist pneumothorax classification sensitivity, it is limited by the number of samples and participants. A more thorough assessment of the effect of saliency maps on the diagnostic performance of radiologists requires more study participants and a structured reading to mitigate confirmation biases. This would also allow for assessment saliency map usefulness for different levels of radiologist expertise. Fourth, following the protocol by Dosovitiskiy et al (8), radiographs were resized to 224 x 224 pixels, making subtle pneumothoraces potentially undetectable.

In conclusion, we investigated the possible use of saliency maps based on ViTs for CXR classification. We showed that ViTs achieved similar results compared with conventional CNNs. The attention-based saliency map, TMME, performed better than the baseline GradCAM, supporting the use of ViTs for CXR classification. Radiologists rated attention-based saliency maps useful in more cases than GradCAM. We showed saliency maps improve model understanding for both AI developers and radiologists, aiding in detection of model biases.

# Acknowledgments

The research for this article received funding from the German federal ministry of health's program for digital innovations for the improvement of patient-centered care in healthcare [grant agreement no. 2520DAT920].

# Tables

**Table 1:** Distribution of Test Images for the User Study

| Model Prediction / Saliency Map | True Positives | False Negatives | True Negatives | False Positives |
|---|---|---|---|---|
| GradCAM | 30 | 15 | 10 | 15 |
| TMME | 30 | 15 | 10 | 15 |
| Artificial | 10 | 10 | 0 | 0 |

Note.—GradCAM = gradient-weighted class activation mapping, TMME = transformer multi-modal explainability

**Table 2:** Comparison of Vision Transformer (ViT) and DenseNet Performance for Pneumothorax Classification on Chest Radiographs

| Data Set | Model | F1 Score | Sensitivity | Specificity |
|---|---|---|---|---|
| **SIIM-ACR** | | | | |
| | ViT | ***66 %** [65 % - 68 %] | ***83 %** [82 %-85 %] | 81 % [80 %-82 %] |
| | DenseNet | 64 % [63 %- 66 %] | 75 % [71 %-78 %] | ***84 %** [82 %-86 %] |
| **Chest X-Ray 14** | | | | |
| | ViT | ***67 %** [65 %-68 %] | ***71 %** [68 %- 73 %] | ***95 %** [95 %-96 %] |
| | DenseNet | 44 % [43 %-46 %] | 55 % [50 %- 63 %] | 89 % [86 %-91 %] |
| **VinBigData** | | | | |
| | ViT | ***40 %** [26 %-56 %] | ***26 %** [13 %-40 %] | **100 %** [100 %-100 %] |
| | DenseNet | 19 % [11 %-33 %] | 19 % [5 %-30 %] | 100 % [98 %-100 %] |

Note.— Maximized F1-scores and the resulting sensitivity and specificity. Data in brackets are 95% CIs. Higher values between the two models are marked with *. SIIM-ACR = Society for Imaging Informatics in Medicine-American College of Radiology

**Table 3:** Cumulative Votes Assessing Saliency Map Usefulness for Pneumothorax Detection by the Three Radiologists in the User Study.

| ViT Prediction | Method | Not useful | Neither | Useful |
|---|---|---|---|---|
| **False Negatives** | | | | |
| | **Artificial** | 26 | 0 | 4 |
| | **GradCAM** | 23 | 7 | 15 |
| | **TMME** | ***18** | 5 | ***22** |
| **False Positives** | | | | |
| | **GradCAM** | 29 | 4 | 12 |
| | **TMME** | ***22** | 4 | ***19** |
| **True Negatives** | | | | |
| | **GradCAM** | 20 | 1 | 9 |
| | **TMME** | ***12** | 7 | ***11** |
| **True Positives** | | | | |
| | **Artificial** | ***2** | 9 | 19 |
| | **GradCAM** | 37 | 7 | 46 |
| | **TMME** | 34 | 9 | ***47** |

Note.— Most useful and least not useful votes are marked with *. GradCAM = gradient-weighted class activation mapping, TMME = transformer multi-modal explainability, ViT = vision transformer

# Tables

# Figures

# Supplementary Figures

# Figure Legends

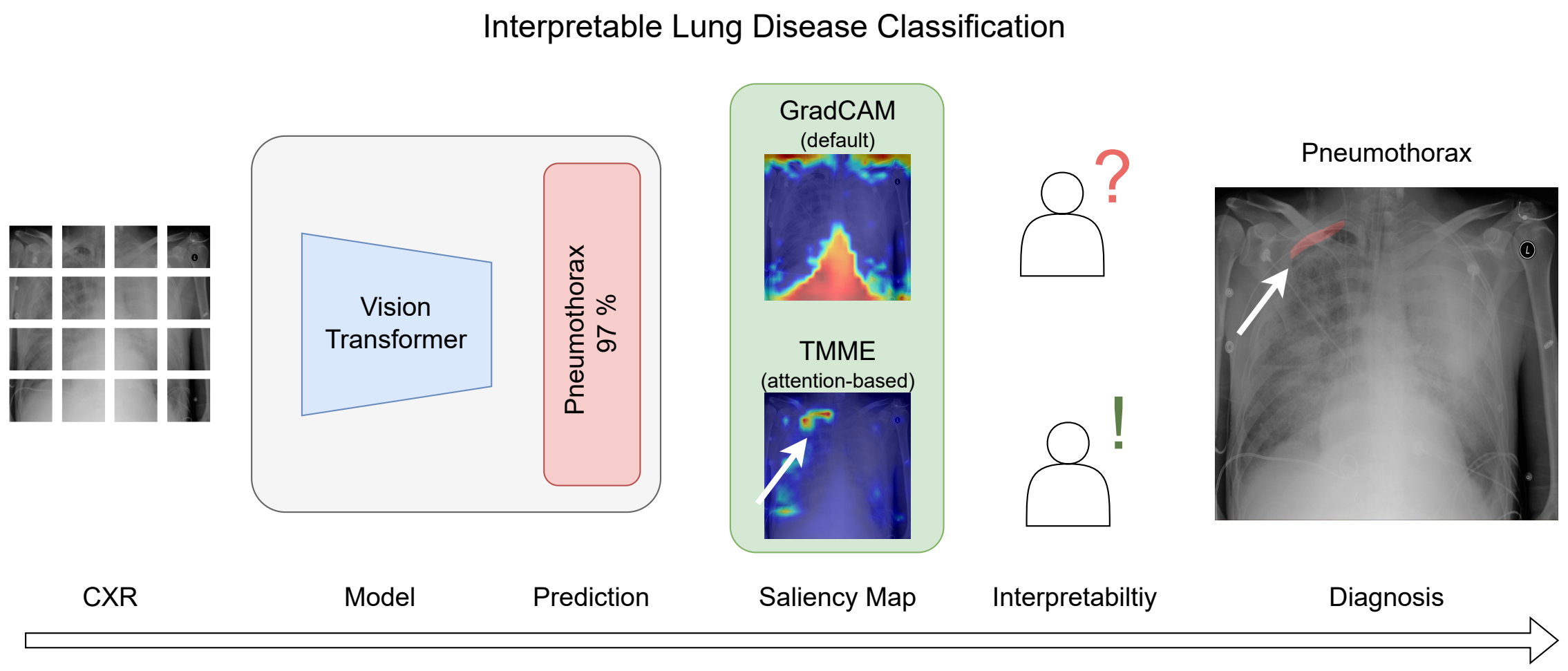

**Figure 1:** Diagram of interpretable lung disease classification pipeline used in this study. After classifying a chest radiograph (CXR) using a vision transformer network, an attention-based saliency map, transformer multi-modal explainability (TMME), is generated to aid the radiologist in interpreting the model's prediction. The proposed saliency map based on TMME was compared with the conventional gradient-weighted class activation map (GradCAM).

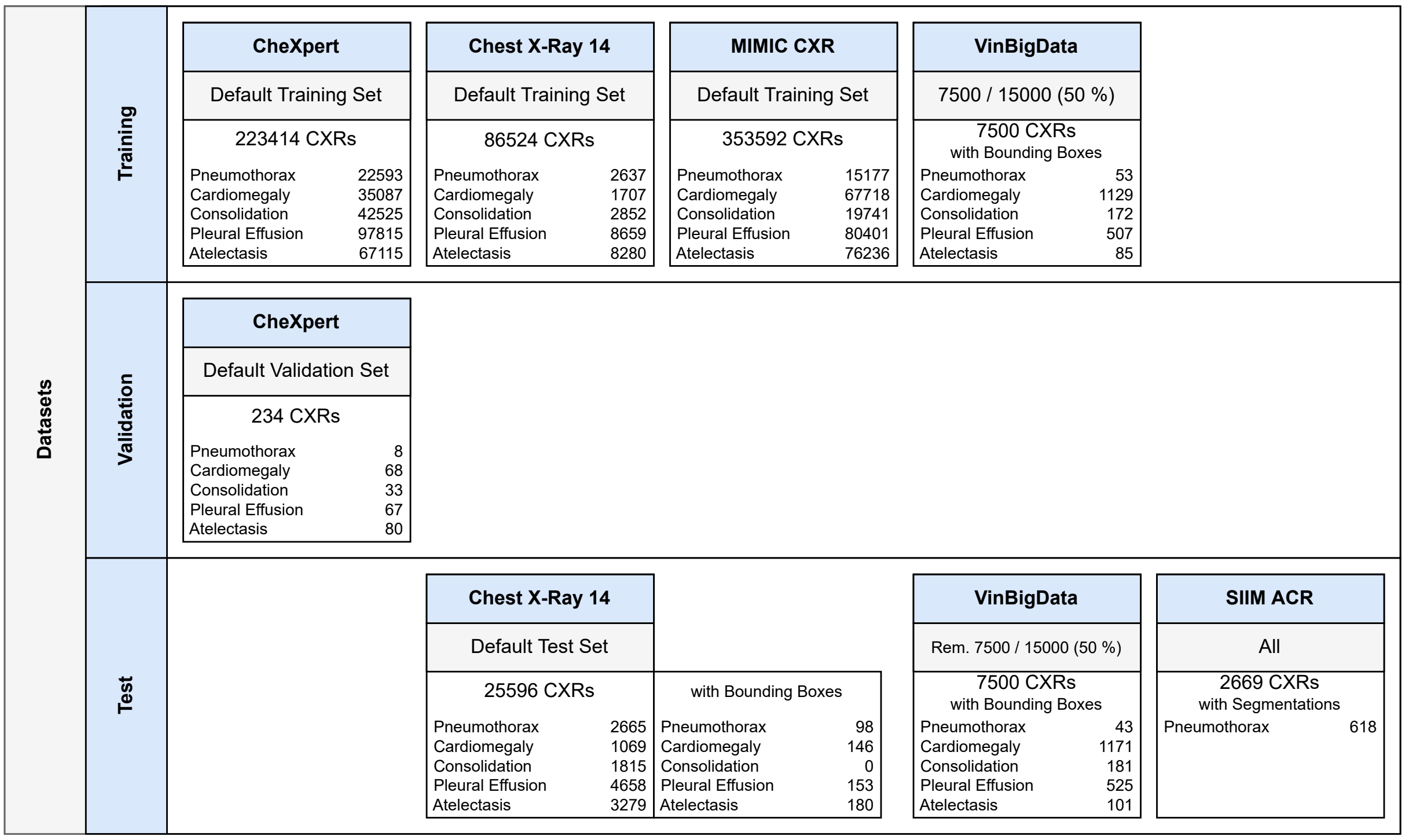

**Figure 2:** Data sets with data splits and classes used in this study. CXR = chest radiograph, SIIM-ACR = Society for Imaging Informatics in Medicine-American College of Radiology

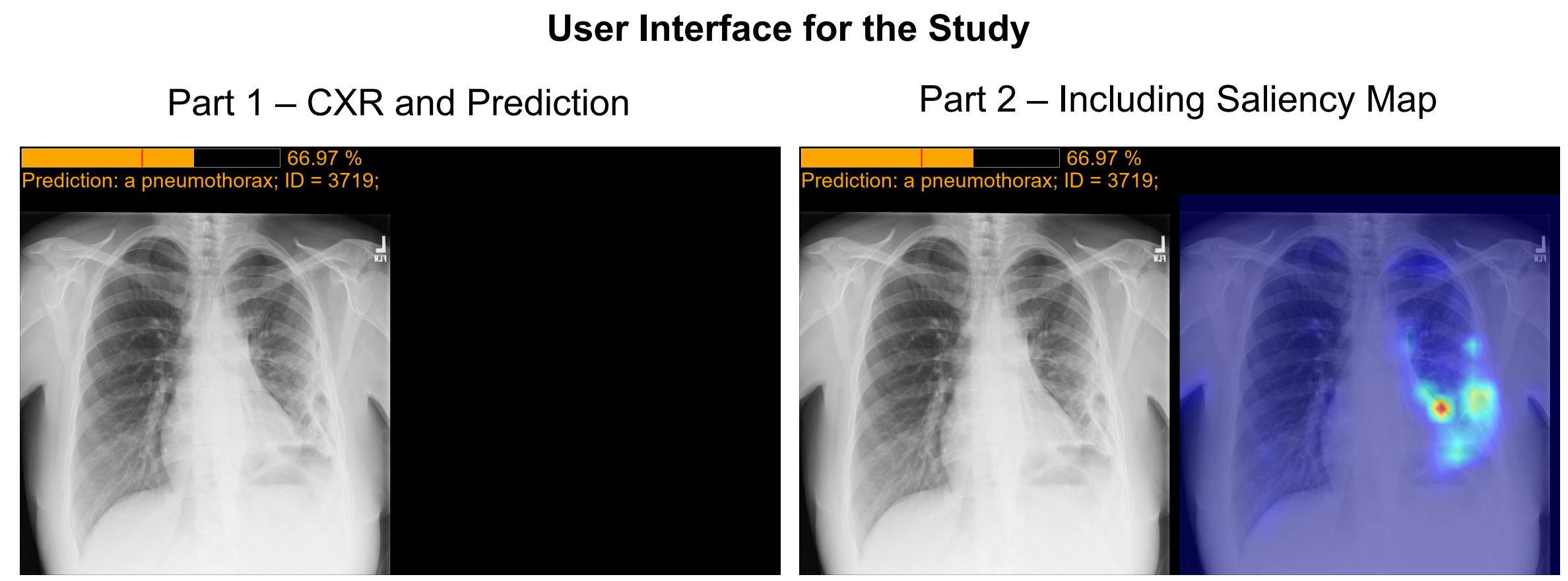

**Figure 3:** User interface for the radiologists in the study. First, they were shown chest radiographs (CXRs) with and without a present pneumothorax and the vision transformer (ViT) prediction score. In the second step, a saliency map was additionally shown. For both parts, radiologists had to detect if a pneumothorax was present and then determine if the saliency map was (subjectively) useful for aiding detection.

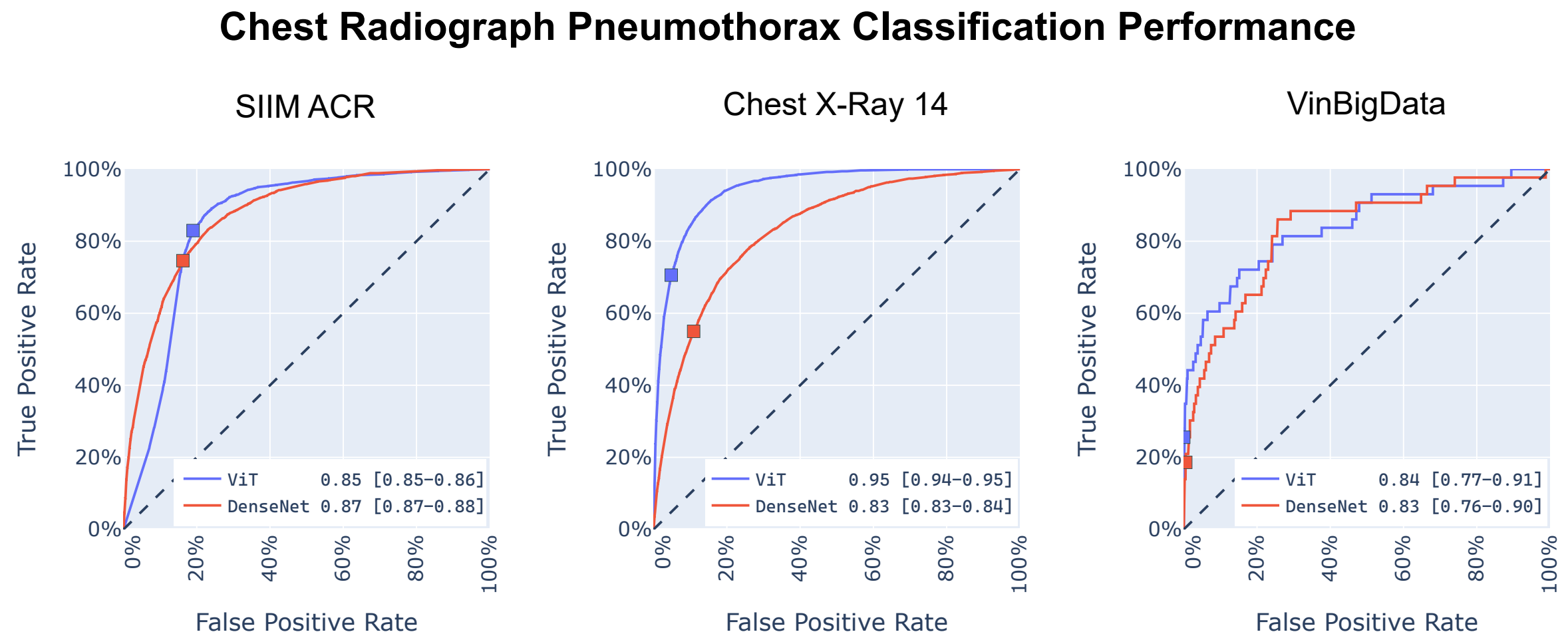

**Figure 4:** Receiver operating characteristic (ROC) curves and areas under the ROC curves [95% CIs] for pneumothorax classification by vision transformer (ViT) and DenseNet models. The ROC curves are computed on hold-out sets. The dots show the maximized F1-scores. SIIM-ACR = Society for Imaging Informatics in Medicine-American College of Radiology

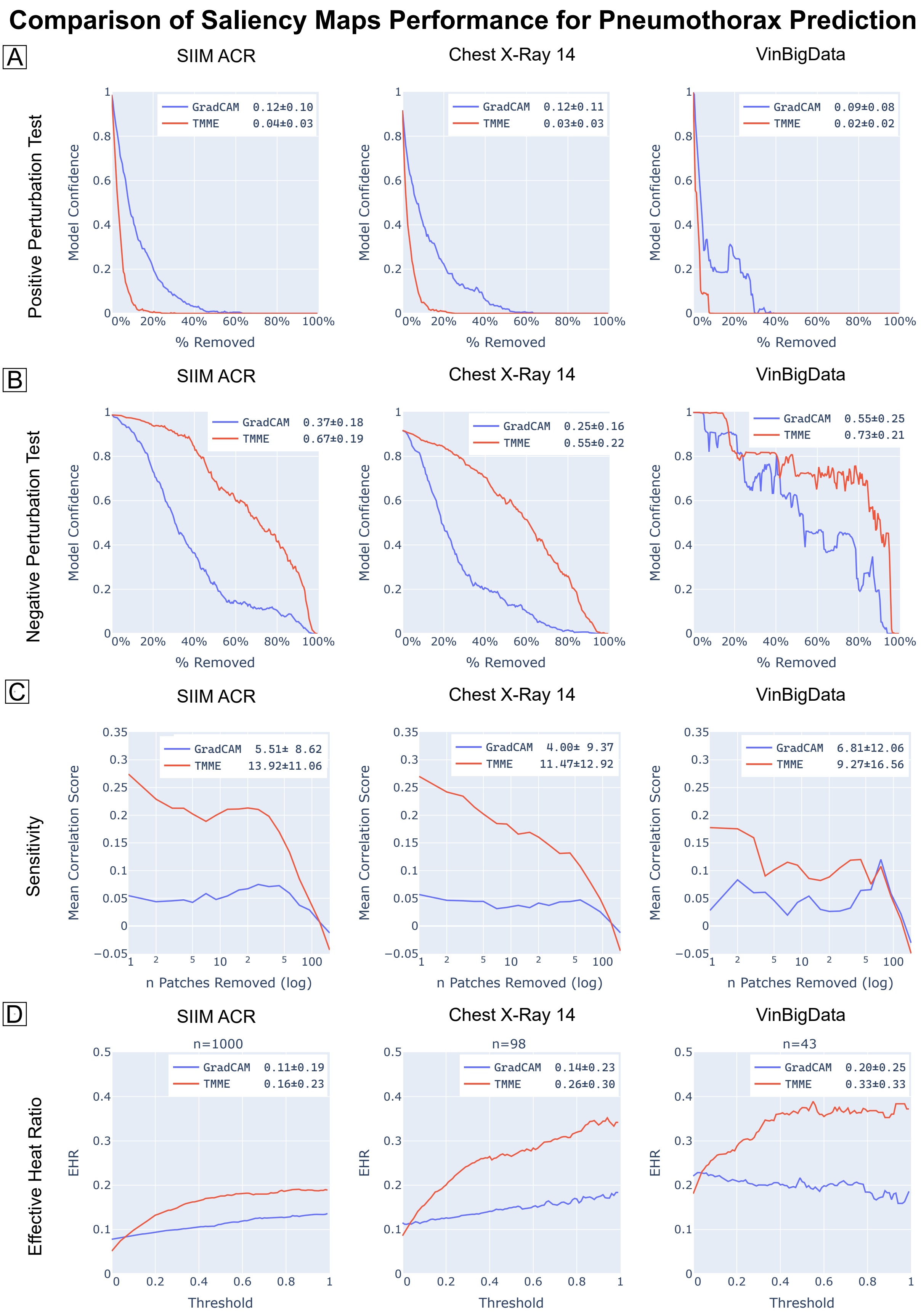

**Figure 5:** Saliency map performance comparison for pneumothorax prediction. Across all tests and data sets transformer multi-modal explainability (TMME) performed significantly

better than gradient-weighted class activation mapping (GradCAM); Saliency metric values ± SD values are the integrals over the respective curves and provided in each graph. **A:** Positive perturbation test; a low value is better. **B:** Negative perturbation test; a high value is better. **C:** Sensitivity-n; a high value is better. **D:** Effective heat ratio (EHR); a high value is better. Results for all classes are shown in **Figure S. 2:** - **Figure S. 5:** EH. SIIM-ACR = Society for Imaging Informatics in Medicine-American College of Radiology

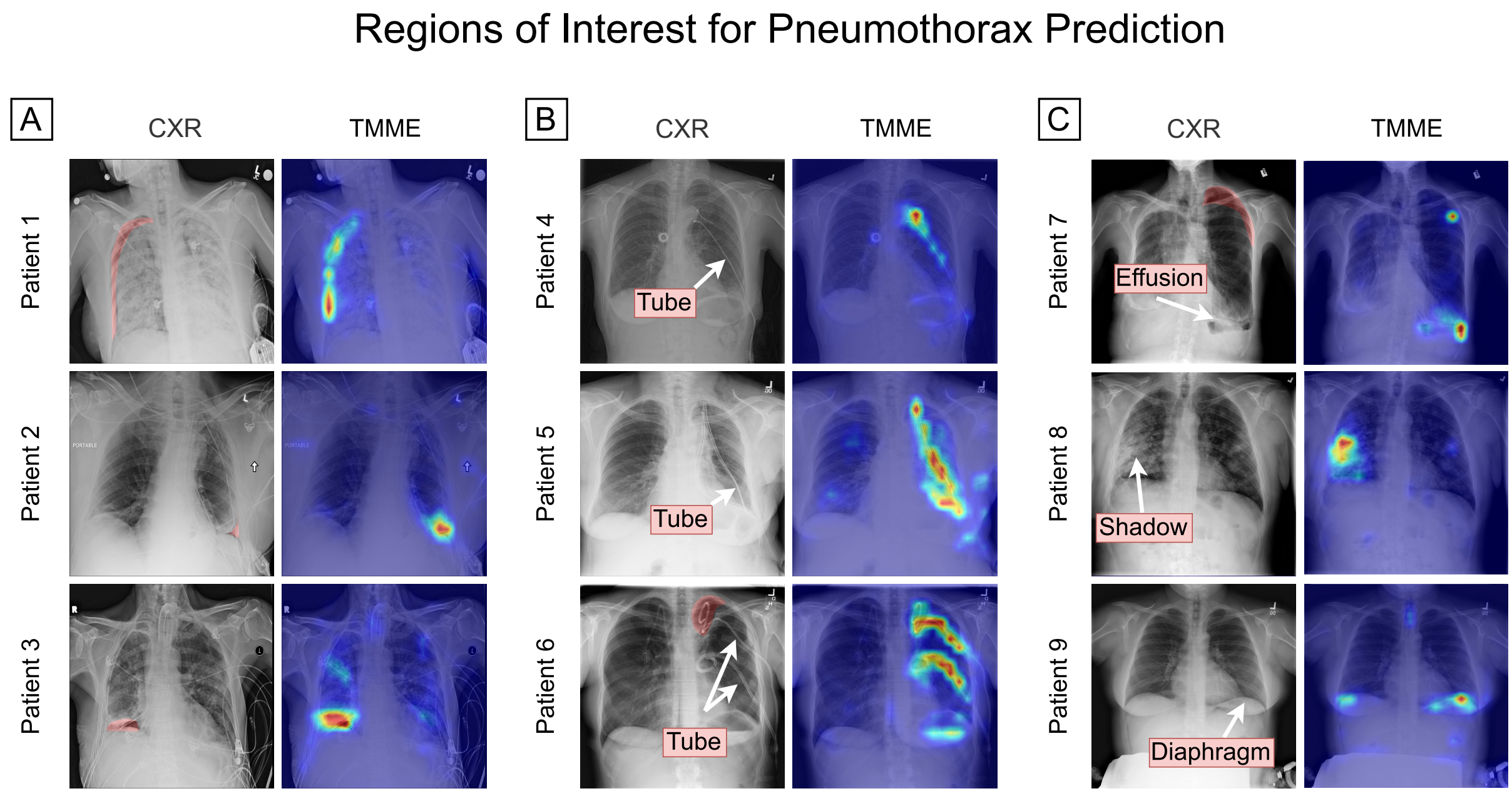

**Figure 6:** Images with vision transformer pneumothorax prediction from SIIM-ACR with transformer multi-modal explainability (TMME) visualization. Ground truth pneumothorax segmentation is highlighted in red. False positives, without pneumothorax, have no inpainting. **A:** Examples where TMME highlights the pneumothorax. **B:** Examples with chest tube highlighting. **C:** Pneumothorax prediction based on other pathologies (plural effusion, lung shadow) or the thoracic diaphragm. CXR = chest radiograph, SIIM-ACR = Society for Imaging Informatics in Medicine-American College of Radiology

# Supplemental Materials

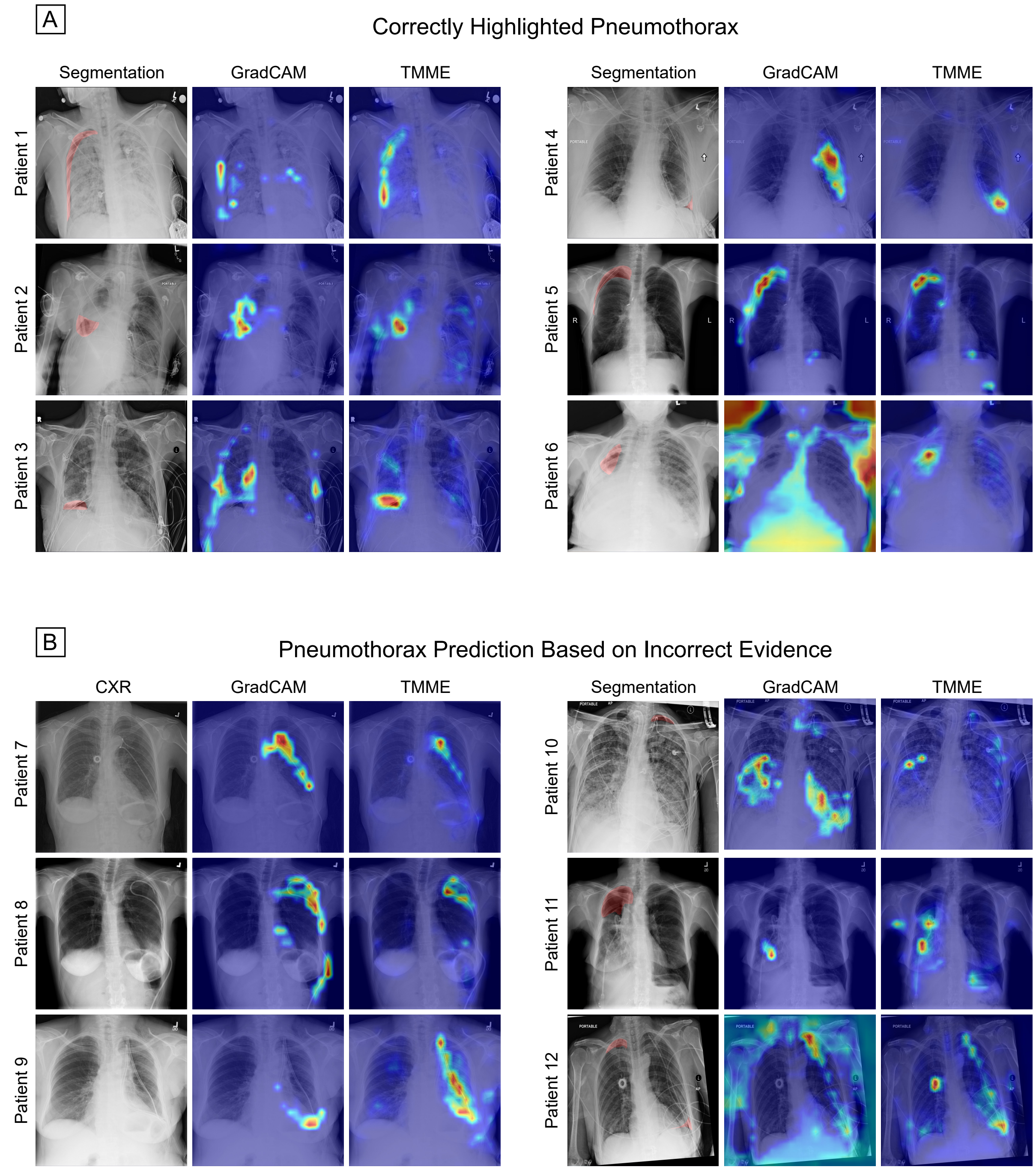

**Figure S. 1:** Saliency maps of pneumothorax predictions based on clinically correct and incorrect evidence. **A**: True positive examples where transformer multi-modal explainability (TMME) highlighted the pneumothorax. **B**: False positives (left) and true positives (right) where the TMME highlights incorrect evidence for pneumothorax. We sampled 50 random true positive and 25 false positive predictions to generate these images.

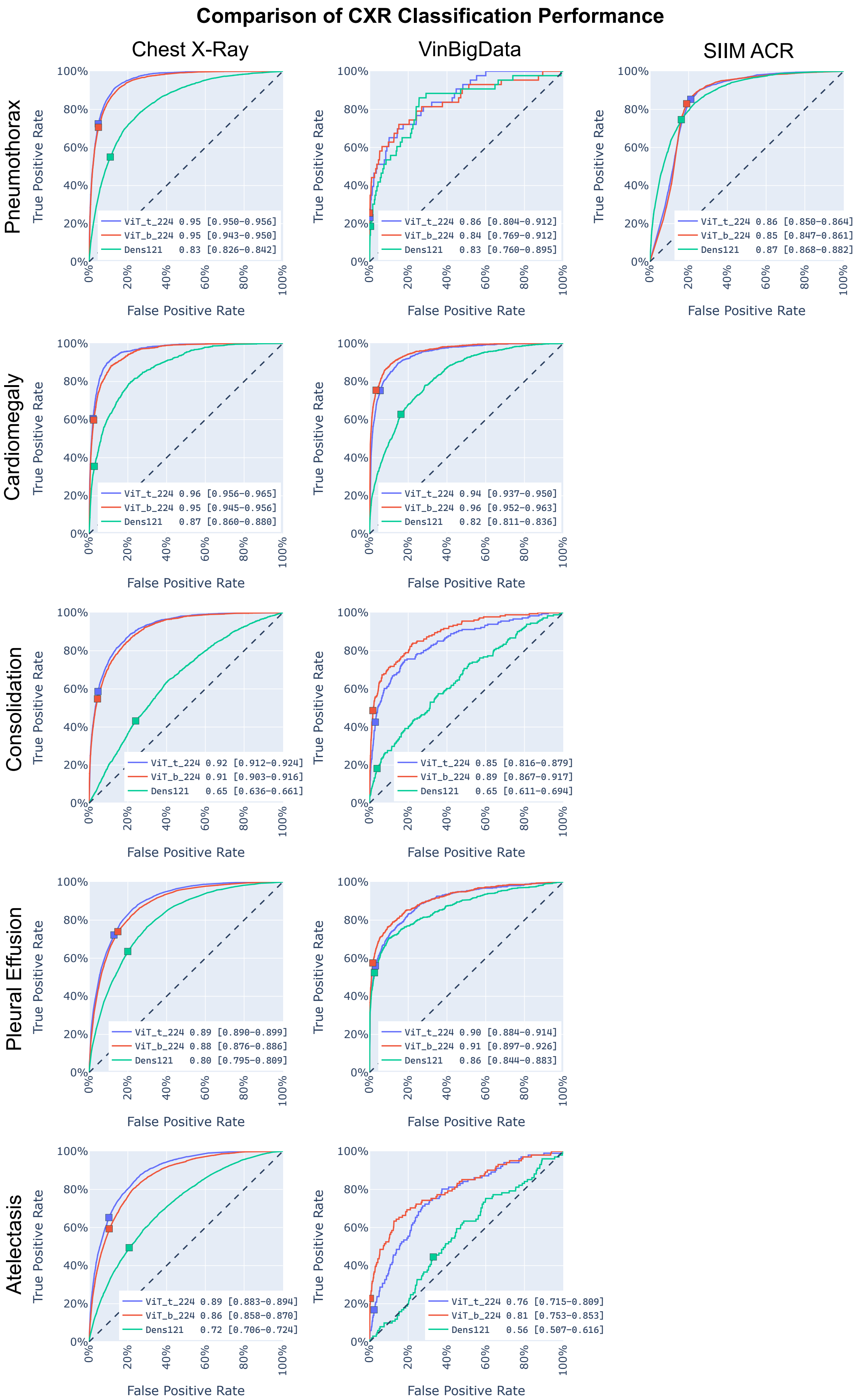

**Figure S. 2:** Receiver operating characteristic curves (ROC) for all five classes: pneumothorax, cardiomegaly, consolidation, pleural effusion, and atelectasis. The SIIM-ACR data set contains only pneumothorax labels. ViT_b_244 (red) is the ViT reported in the main text. During prototyping we additionally tested the smaller version, ViT_t_244 (blue). The area under the ROC curve and 95 % confidence intervals are reported in the legend. SIIM-ACR = Society for Imaging Informatics in Medicine-American College of Radiology.

## Comparison of Saliency Maps: Negative Pertubation

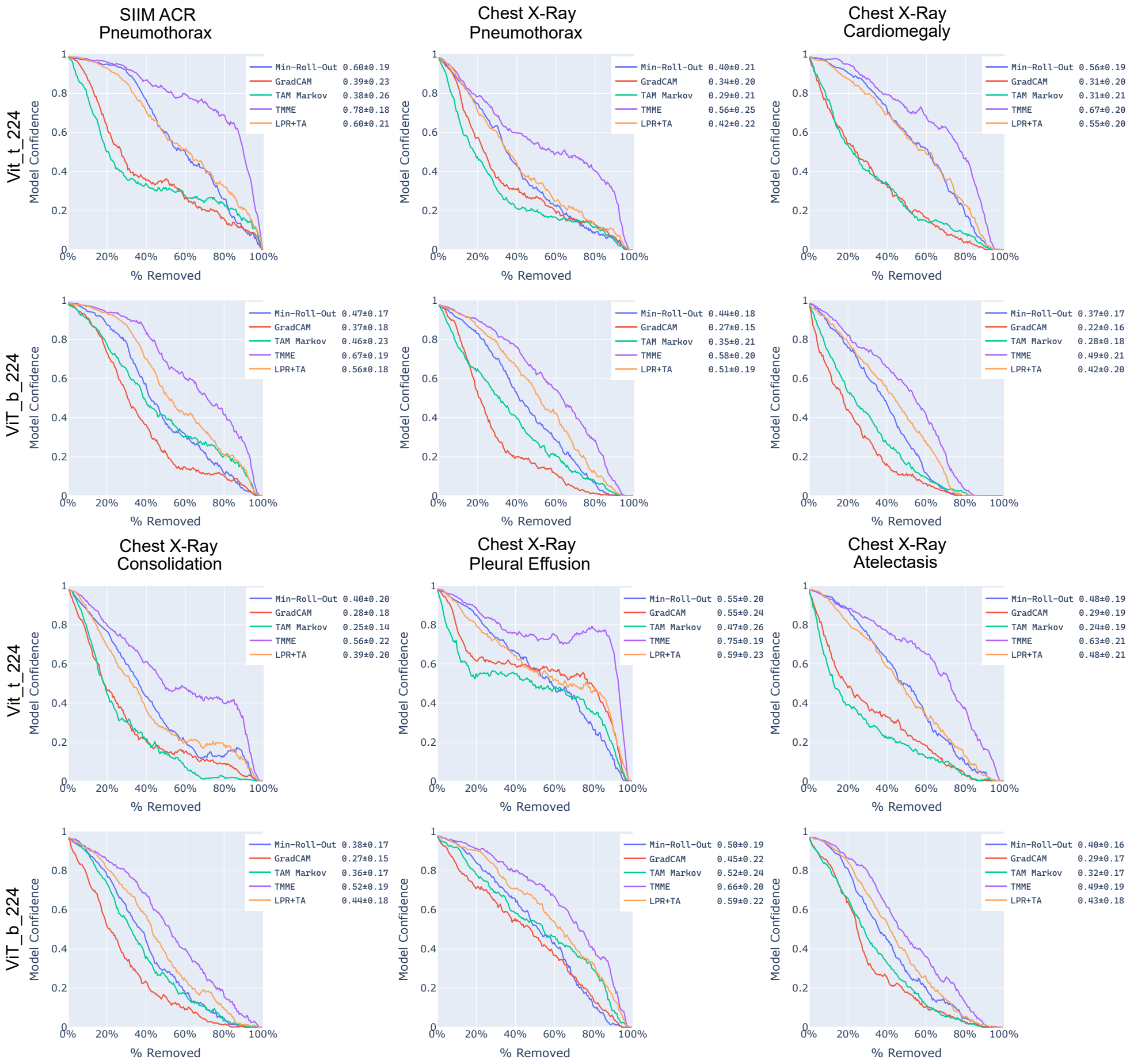

## Comparison of Saliency Maps: Positive Pertubation

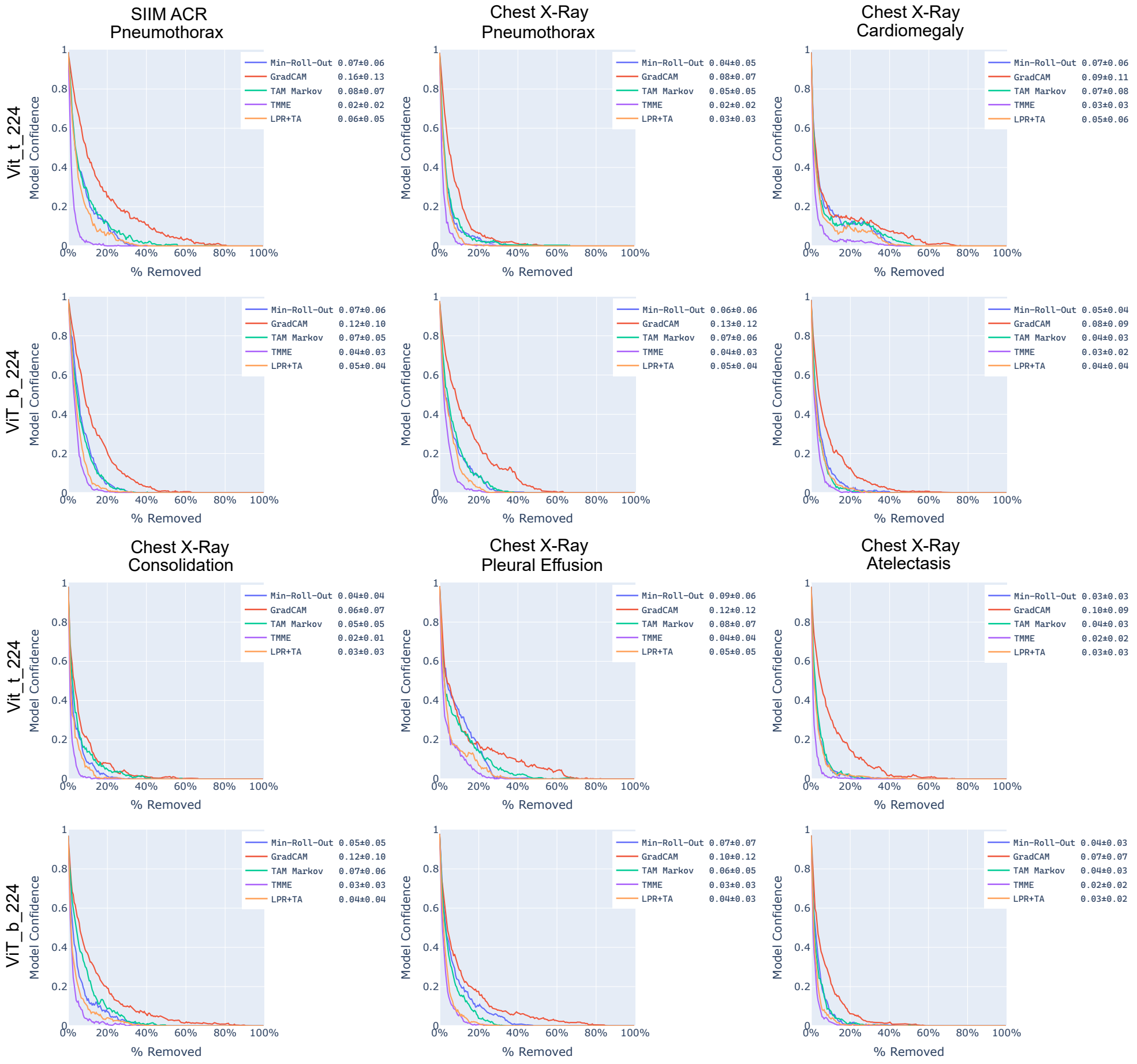

**Figure S. 3:** Positive and negative perturbation test on all 5 classes: pneumothorax, cardiomegaly, consolidation, pleural effusion, and atelectasis. ViT_b_244 is the same ViT as reported in the text, ViT_t_244 is the smaller version used during prototyping. During prototyping we investigated other, less performing, attention-based saliency map methods: min-roll-out, transition attention maps (TAM) Markov, and layer-wise relevance propagation transformer-attribution (LRP+TA). The area under the receiver operating characteristic curves and SD are reported in the legend. The results are averaged over 250 true positive examples.

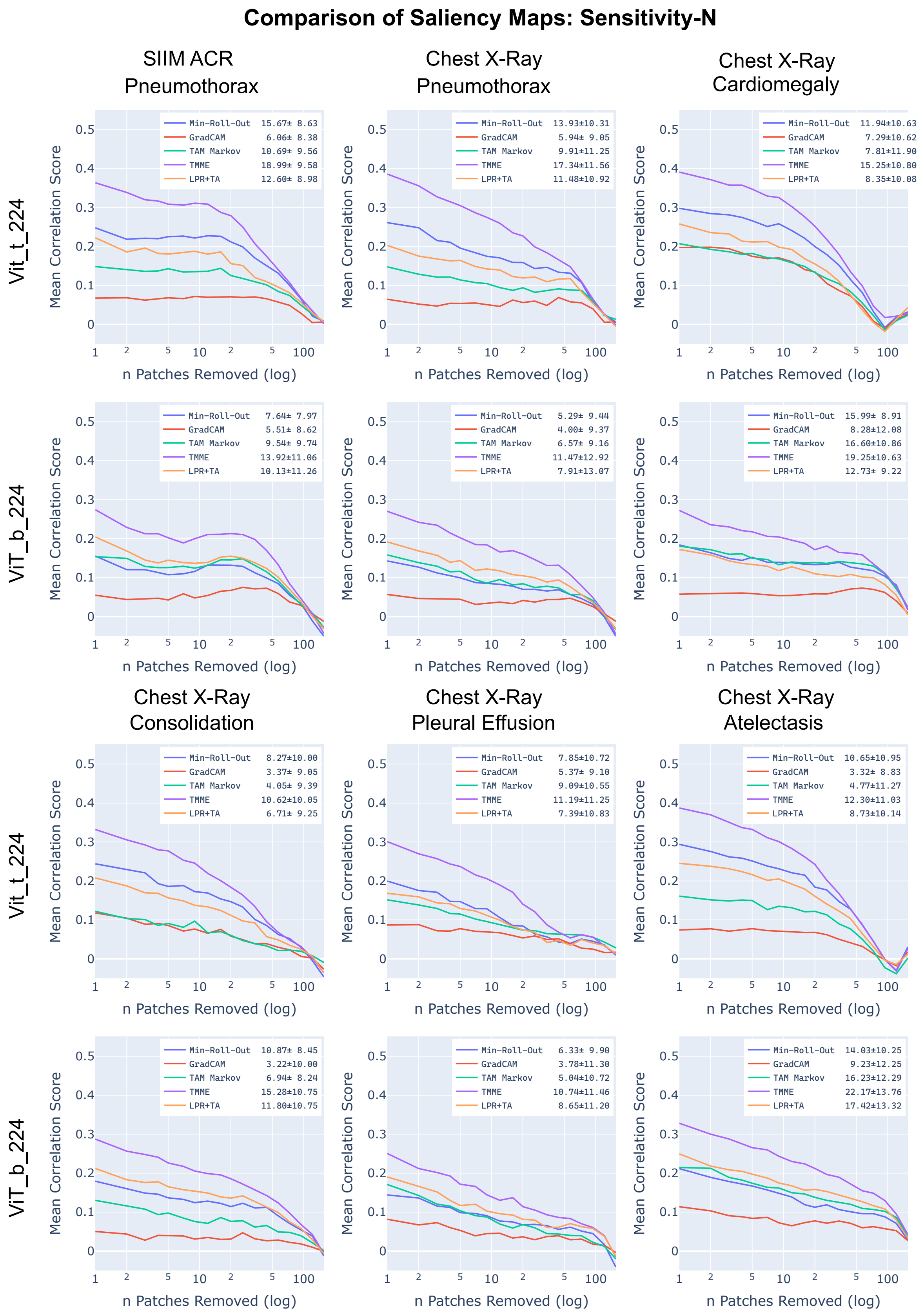

**Figure S. 4:** Sensitivity-n test on all 5 classes: pneumothorax, cardiomegaly, consolidation, pleural effusion, and atelectasis. ViT_b_244 is same ViT as reported in the text, ViT_t_244 is the smaller version used during prototyping. During prototyping we investigated other, less performing, attention-based saliency map methods: min-roll-out, transition attention maps (TAM) Markov, and layer-wise relevance propagation transformer-attribution (LRP+TA). The area under the receiver operating characteristic curves and SD are reported in the legend.

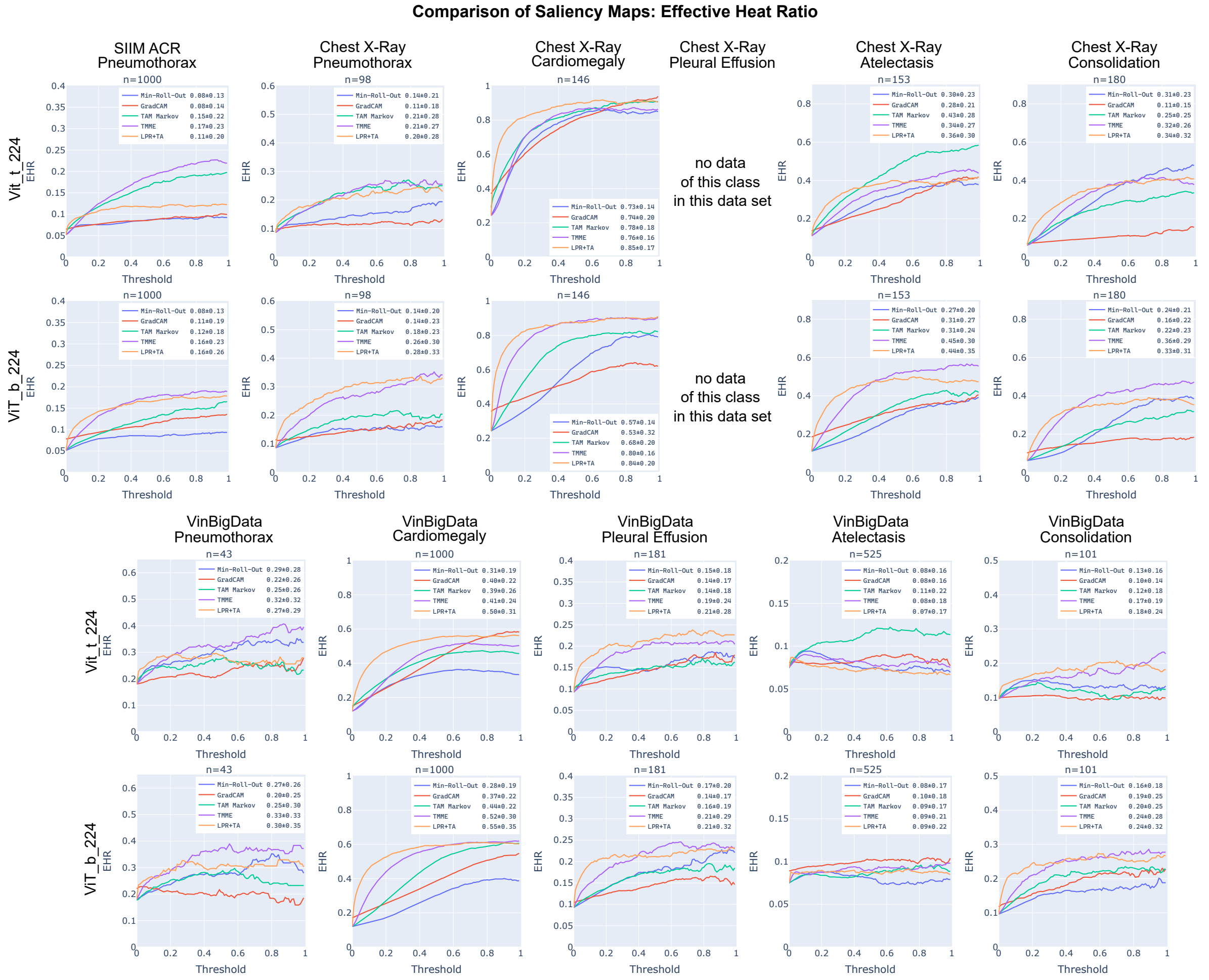

**Figure S. 5:** EHR test on all 5 classes: pneumothorax, cardiomegaly, consolidation, pleural effusion, and atelectasis. ViT_b_244 is same ViT as reported in the text. While ViT_t_244 is the smaller version used during prototyping. During prototyping we investigated other, less performing, attention-based saliency map methods: min-roll-out, transition attention maps (TAM) Markov, and layer-wise relevance propagation transformer-attribution (LRP+TA). The area under the receiver operating characteristic curves and SD are reported in the legend. The results are averaged over n segmented examples, where n is reported in the diagram.

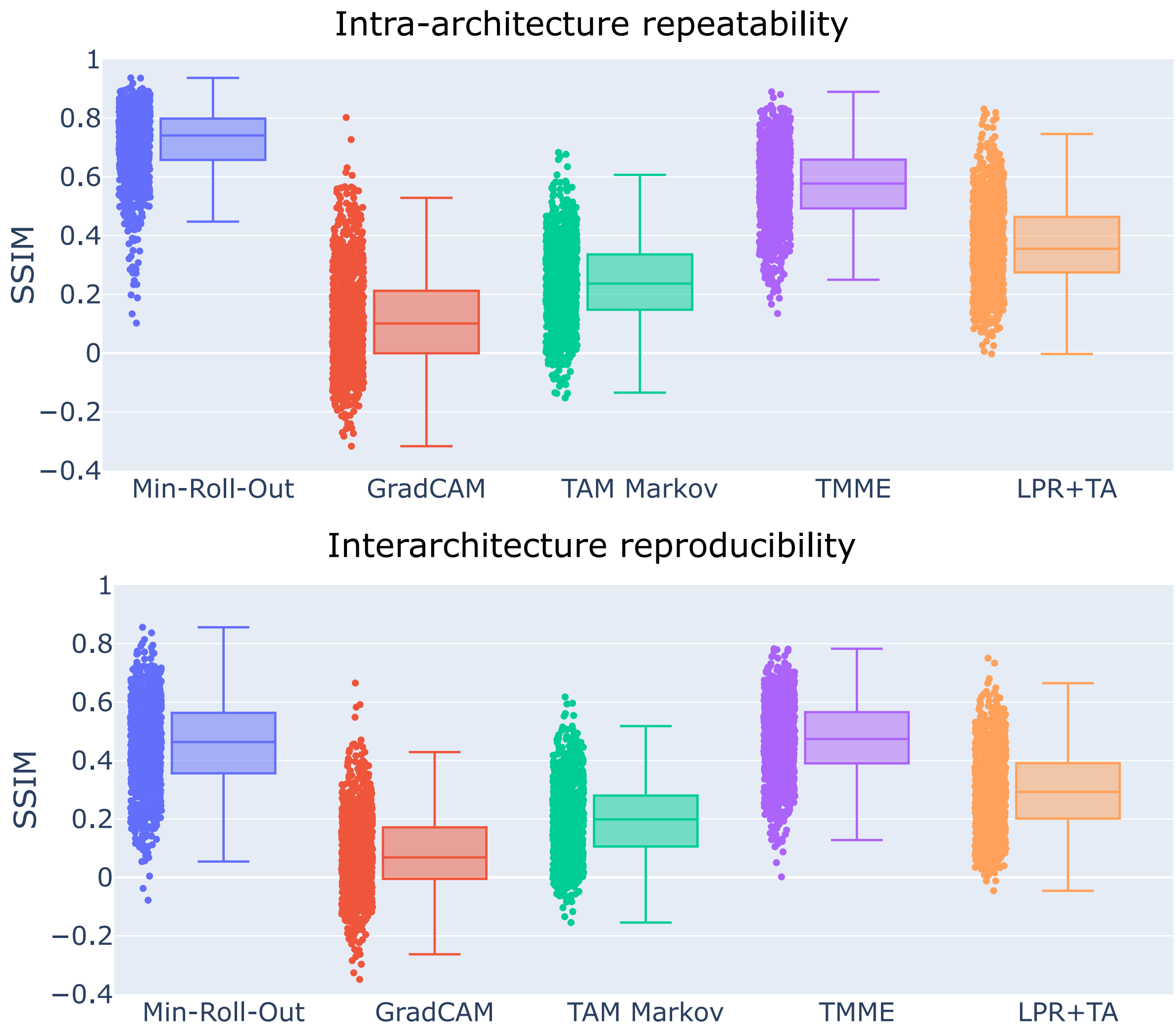

**Figure S. 6:** Intra-architecture repeatability and interarchitecture reproducibility.

Intra-architecture repeatability is the SSIM score of two saliency maps of the same ViT (ViT_b_244) from different training sessions. Min-Roll-Out 0.71 (95 % CI: 0.71, 0.72); GradCAM 0.116 (95 % CI: 0.11, 0.13); TAM Markov 0.24 (95 % CI: 0.23, 0.25); TMME 0.57 (95 % CI: 0.56, 0.58); LPR+TA 0.37 (95 % CI: 0.36, 0.38); interarchitecture reproducibility is the SSIM score of two saliency maps of two different architectures: the reported ViT (ViT_b_244) and the smaller ViT (ViT_t_244). Min-roll-out 0.46 (95 % CI: 0.45, 0.47); GradCAM 0.082 (95 % CI: 0.074, 0.091); TAM Markov 0.20 (95 % CI: 0.19, 0.21); TMME 0.47 (95 % CI: 0.47, 0.48); LPR+TA 0.30- (95 % CI: 0.29, 0.31). During prototyping we investigated other, less performing, attention-based saliency map methods: min-roll-out, transition attention maps (TAM) Markov, and layer-wise relevance propagation transformer-attribution (LRP+TA). SSIM = structured similarity index measure.

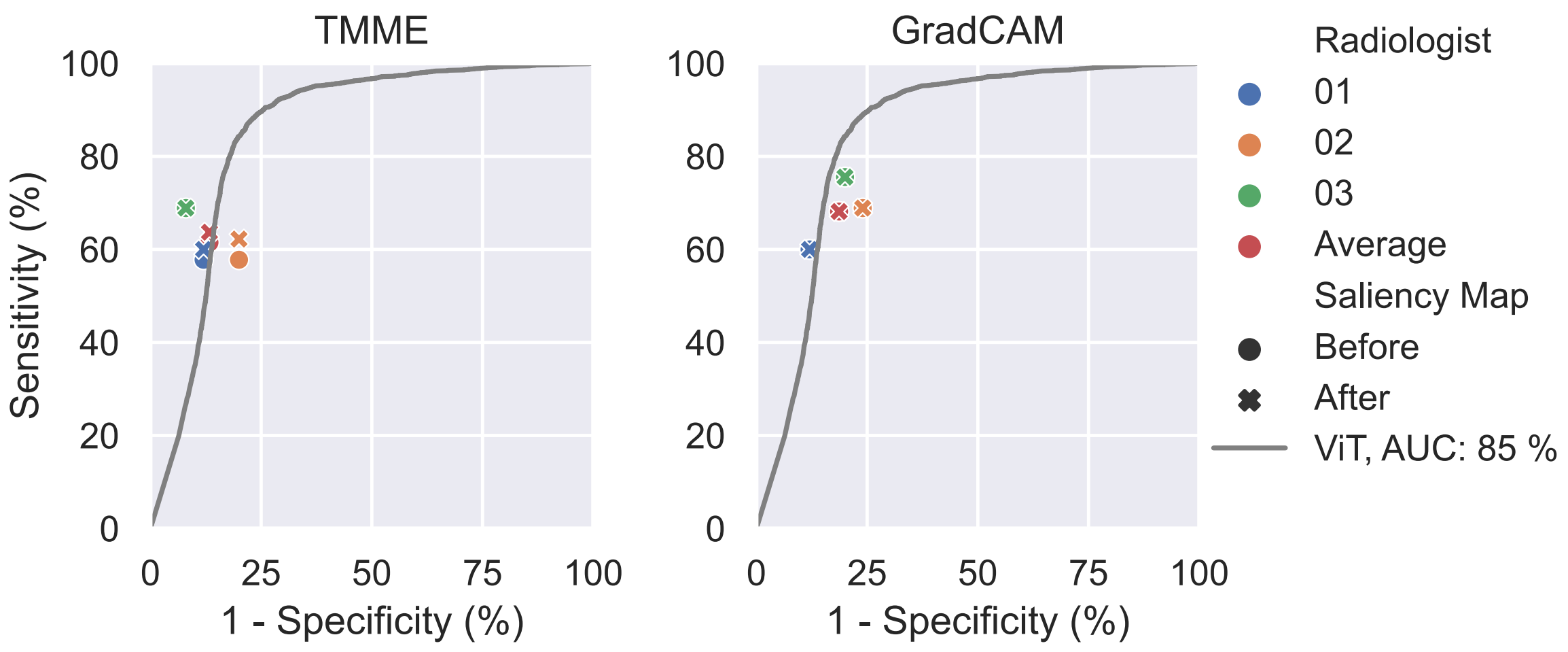

**Figure S. 7:** Receiver operating characteristic curves (ROC) and model AUC for the user study. Performance of the ViT model (grey) and the radiologists without (circles) and with (crosses) additional saliency map for pneumothorax classification. The average reader performance is shown by the red circle/cross. AUC = area under the ROC curve.

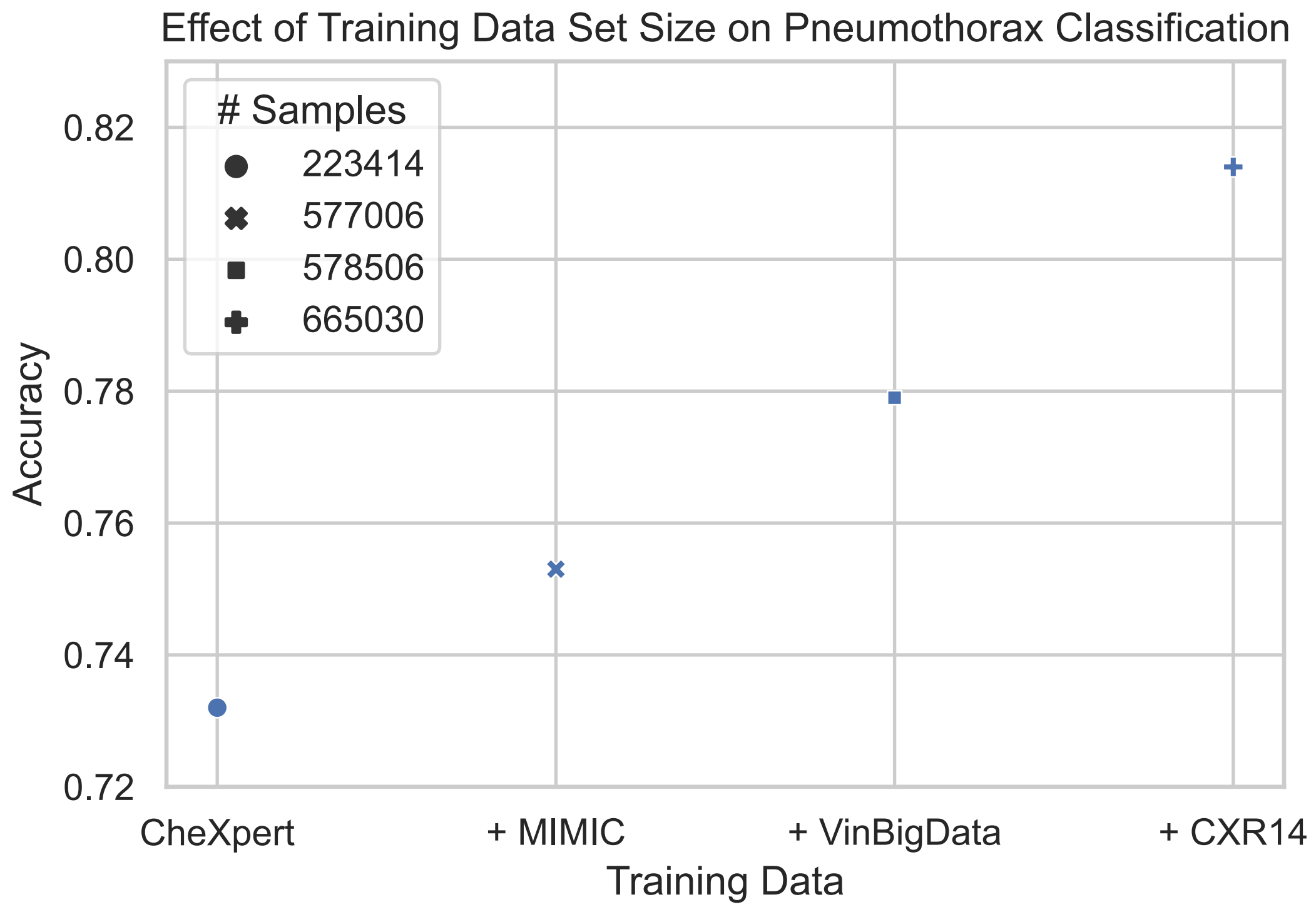

**Figure S. 8**: Effect of increased training data on pneumothorax classification on the SIIM ACR test data set. The classification thresholds were computed according to the best respective F1-score on the validation data. All networks were trained for the same number of iterations. CXR14 = Chest X-Ray 14, SIIM-ACR = Society for Imaging Informatics in Medicine-American College of Radiology.